\documentclass[runningheads]{llncs}
\usepackage{graphicx}
\usepackage{multirow}
\usepackage{soul,color}
\usepackage[table]{xcolor}
\usepackage{booktabs}

\begin{document}
\title{Multi-Feature Semi-Supervised Learning for COVID-19 Diagnosis from Chest X-ray Images}

\author{
	Xiao Qi \inst{1}   \and
	John L. Nosher\inst{2}     \and
	David J. Foran \inst{3}    \and
	Ilker Hacihaliloglu\inst{2,4}
}
\authorrunning{Xiao Qi et al.}
\institute{
	Department of Electrical and Computer Engineering, Rutgers University, Piscataway, NJ, USA   \and
	Department Radiology, Rutgers Robert Wood Johnson Medical School, New Brunswick, NJ, USA     \and
	Rutgers Cancer Institute of New Jersey, New Brunswick, NJ, USA             \and 
	Department Biomedical Engineering, Rutgers University, Piscataway, NJ, USA    
}

%\institute{Princeton University, Princeton NJ 08544, USA \and
%Springer Heidelberg, Tiergartenstr. 17, 69121 Heidelberg, Germany
\maketitle              % typeset the header of the contribution
\begin{abstract}
Computed tomography (CT) and chest X-ray (CXR) have been the two dominant imaging modalities deployed for improved management of Coronavirus disease 2019 (COVID-19). Due to faster imaging, less radiation exposure, and being cost-effective CXR is preferred over CT. However, the interpretation of CXR images, compared to CT, is more challenging due to low image resolution and COVID-19 image features being similar to regular pneumonia. Computer-aided diagnosis via deep learning has been investigated to help mitigate these problems and help clinicians during the decision-making process. The requirement for a large amount of labeled data is one of the major problems of deep learning methods when deployed in the medical domain. To provide a solution to this, in this work, we propose a semi-supervised learning (SSL) approach using minimal data for training. We integrate local-phase CXR image features into a multi-feature convolutional neural network architecture where the training of SSL method is obtained with a teacher/student paradigm. Quantitative evaluation is performed on 8,851 normal (healthy), 6,045 pneumonia, and 3,795 COVID-19 CXR scans. By only using 7.06\% labeled and 16.48\% unlabeled data for training, 5.53\% for validation, our method achieves 93.61\% mean accuracy on a large-scale (70.93\%) test data. We provide comparison results against fully supervised and SSL methods. The code can be accessed from \url{https://github.com/endiqq/Multi-Feature-Semi-Supervised-Learning-for-COVID-19-CXR-Images}
%The abstract should briefly summarize the contents of the paper in 150--250 words.

\end{abstract}
\section{Introduction}

Diagnostics is a key tool for improved management of Coronavirus disease 2019 (COVID-19), permitting healthcare workers to rapidly triage patients. Currently, the gold standard diagnosis is based on reverse-transcription polymerase chain reaction (RT-PCR) tests. To improve the management of COVID-19, radiological assessment, based on computed tomography (CT) and chest X-ray (CXR), has also been incorporated into the decision-making process. Compared to CT, CXR provides additional advantages such as fast screening, being portable, and easy to setup (can be setup in isolation rooms). However, the interpretation of CXR images, compared to CT, by the expert radiologist is a difficult process as the visual cues for the disease can be subtle or similar to regular pneumonia. As such computer-aided diagnostic systems that can aid in the decision-making process have been investigated \cite{ozturk2020automated,qi2020chest,wang2020covid}.

Computer-aided diagnosis via deep supervised learning has achieved strong performance when provided with a large labeled data set \cite{ozturk2020automated,qi2020chest,wang2020covid}. However, the requirement of expert knowledge for the labeling process is costly. The scarcity of available data, in particular with a new disease, also affects the success of the developed methods. In order to mitigate this problem methods based on semi-supervised learning (SSL) have been developed for diagnosing lung disease from CXR images. 

The following papers provide a brief overview of prior work on lung disease classification from CXR data using SSL. Consistency regularization and pseudo labeling have been the two most dominant methods investigated for SSL. \cite{unnikrishnan2020semi} proposed a multi-class abnormality detection from CXR data. Using 35\% labeled and 35\% unlabeled data as the training set an AUROC close to 0.82 was reported on 20\% test data. In \cite{gyawali2019semi}, disentangled stochastic latent space was used to improve self-ensembling for semi-supervised CXR classification. Using 3.12\% labeled and 73.2\% unlabeled data for training an AUROC value of 0.66 was obtained on 22\% test data. In \cite{gyawali2020semi}, the same group extended their prior work \cite{gyawali2019semi} by training SSL network on linear mixing of labeled and unlabeled data, at the input and latent space, to improve network regularization. 9.6\% improvement of AUROC, compared to \cite{gyawali2019semi}, was reported using same evaluation strategy as \cite{gyawali2019semi}. In \cite{aviles2019graphx}, the authors proposed a graph-based label propagation method for semi-supervised CXR classification and achieved an AUROC of 0.78 with only using 20\% of the labeled data. Unlabeled dataset size used for training and test dataset size was not reported in \cite{aviles2019graphx}. \cite{aviles2019graphx} was recently extended for SSL-based classification of Covid-19 from CXR data \cite{aviles2020graphxcovid}. Using 30\% labeled and 70\% unlabeled data for training an average accuracy of 94.6\% was reported on 2.14\% test data. Training and testing data included only 200 and 100 Covid-19 scans respectively.

Deep learning methods can automatically learn the features of the data without the need for data preprocessing. Nonetheless, automation requires 1- the development of deeper and complex networks, and 2- annotated large training data. Preprocessing can provide solutions to some of these challenges. Most recently \cite{qi2020chest} local phase-based image processing was proposed for improved representation of lung CXR data. Quantitative results demonstrated the importance of local phase image features for improved diagnosis of COVID-19 disease from CXR scans. Motivated by this, in this work we propose a local phase-based SSL method for accurate diagnosis of COVID-19 from minimal training data. Our proposed method achieves similar accuracy as full supervised baseline architecture accuracy by only using 7.06\% labeled and 16.48\% unlabeled data for training, 5.53\% for validation, on 70.93\% testing data. Our contributions are as follows: 1- We introduce a novel multi-feature SSL model which outperforms baseline mono-feature-based SSL using minimal training data. 2- We perform ablation studies to show the effect of local-phase features for training and testing. 3- We evaluate our technique on 18,691 CXR data set of healthy (8,851), regular pneumonia (6,045), and COVID-19 (3,795) scans. This is the largest COVID-19 evaluation study reported for SSL. We provide a performance comparison of the presented method against the supervised baseline and several SSL methods.

\section{Methods and Materials}
\begin{figure}
	\begin{center}
		\includegraphics[width=9cm]{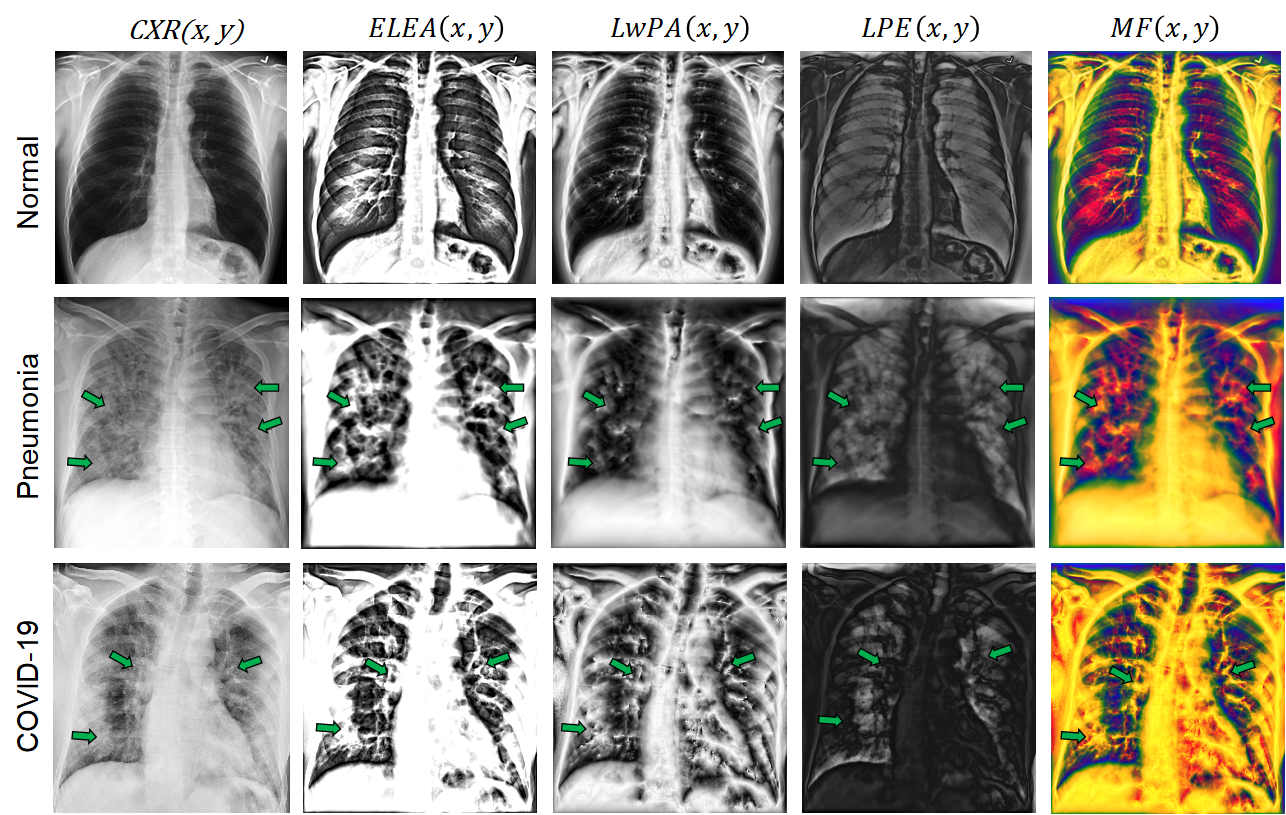}
		\caption{$CXR(x,y)$ from normal, bacterial pneumonia and COVID-19 lung and corresponding enhanced images. Green arrows point to consolidations.}
		\label{fig:enhance}
	\end{center}
\end{figure}

\noindent\textbf{Multi-feature Images:} Enhancement of CXR images ($CXR(x,y)$) is based on the extraction of local phase image features using bandpass quadrature filters and L1 norm-based contextual regularization method \cite{qi2020chest}. Monogenic filter, and $\alpha$-scale space derivative quadrature filters (ASSD) are used as bandpass quadrature filters. Three different local phase $CXR(x,y)$ image features are extracted: 1-  Local weighted mean phase angle ($LwPA(x,y)$), 2- $LwPA(x,y)$ weighted local phase energy ($LPE(x,y)$), and 3- Enhanced local energy attenuation image ($ELEA(x,y)$). During this work, we used the same filter parameters as explained in \cite{qi2020chest}. A multi-feature image, denoted as $MF(x,y)$, is created by combining these three types of local phase images as three-channel input. Since all enhanced images are grayscale, the combination results in an image with shape (\textit{w*h*3}), where $w$ and $h$ correspond to the width and height of the image. Qualitative results corresponding to the $MF(x,y)$ images are displayed in Figure \ref{fig:enhance}. Green arrows point to diffuse irregular patchy consolidations. The COVID-19 image shows bilateral peripherally distributed opacities (green arrows). Investigating Figure \ref{fig:enhance} we can observe local phase and $MF(x,y)$ images of COVID-19 have improved opacity features related to COVID-19 compared to original $CXR(x,y)$ image. Although $LPE(x,y)$ image did not result in opacity detection improvement in the COVID -19 image, it could compliment the other two local phase enhanced CXR images. We can observe a similar enhancement of consolidations related to bacterial pneumonia image (middle row). The $MF(x,y)$ image and the $CXR(x,y)$ image are used as an input to train our proposed multi-feature teacher/student SSL model which is explained in the next section.

%Investigating Figure \ref{fig:enhance} we can observe that the enhanced local phase images extract new lung features that are not visible in the original $CXR(x,y)$ images. Pneumonia image is obtained from a patient diagnosed with human bocavirus. 

\begin{figure}
	\begin{center}
		\includegraphics[width=11cm]{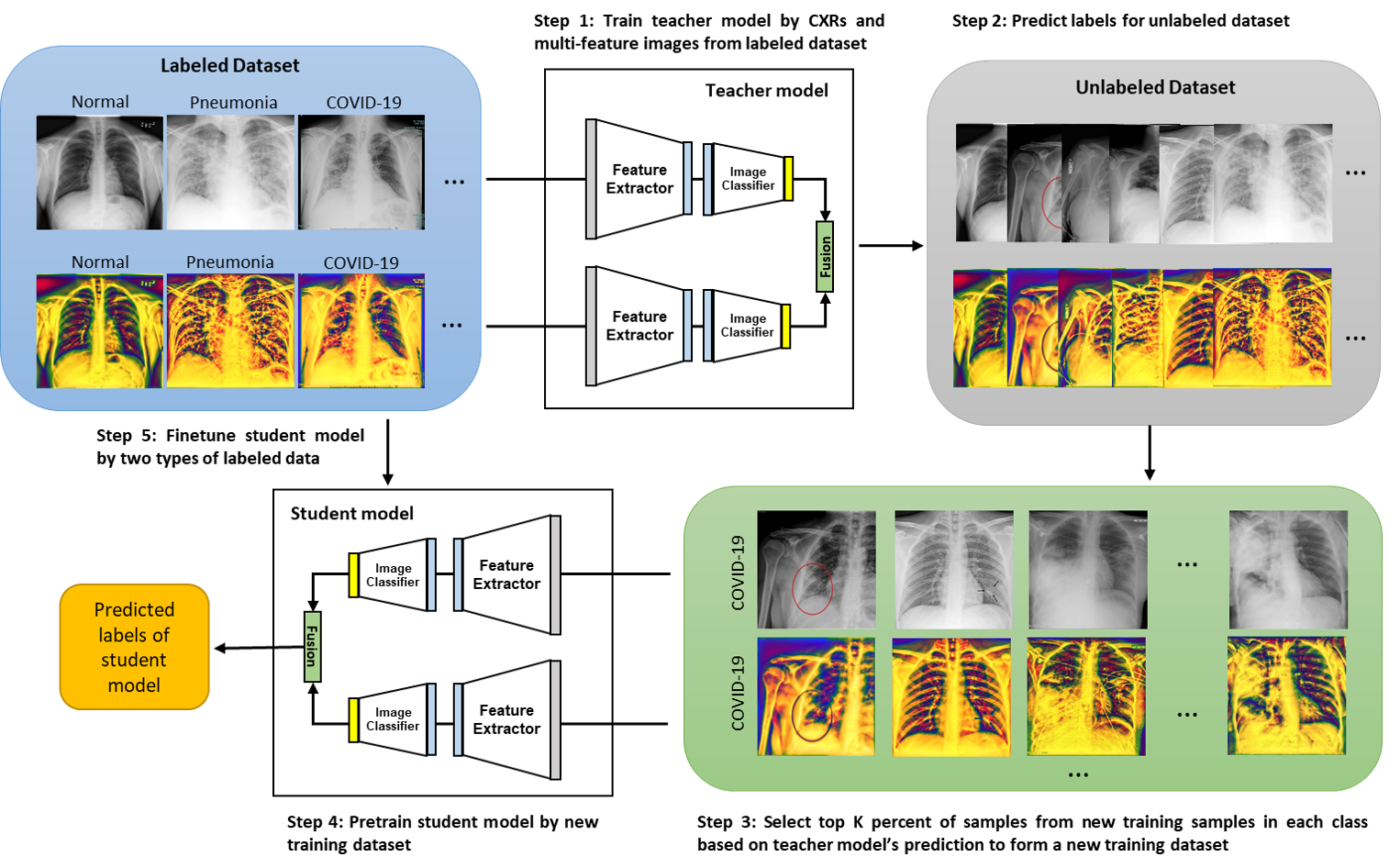}
		\caption{Illustration of our proposed multi-feature guided SSL method. Step 1: Train a multi-feature teacher model on labeled dataset; Step 2: Run the trained multi-feature teacher model on unlabeled dataset to obtain the pseudo-labeled dataset; Step 3: Optimize the pseudo-labeled dataset by using hyperparameter K ; Step 4: Train a new student model on the optimized pseudo-labeled dataset; and Step 5: Finetune the trained student model on the labeled dataset.}
		\label{pipeline}
	\end{center}
\end{figure}

\noindent\textbf{Semi-supervised learning pipeline:} Our proposed method consists of five steps as illustrated in Figure \ref{pipeline}. The pipeline is an extension of \cite{DBLP:journals/corr/abs-1905-00546}, and the algorithm leverages the desirable property that learning is tolerant to a certain degree of noise\cite{10.5555/2999611.2999745}. Our goal here is to use a multi-feature convolutional neural network (CNN) architecture for the teacher model to limit the labeling noise for unlabeled dataset, thus forming a better teacher model without additional training samples. The same multi-feature CNN architecture is also used for the student model.

Our multi-feature CNN architecture consists of two mono-feature network streams for processing $CXR(x,y)$ images and the enhanced $MF(x,y)$ images respectively. In \cite{alsinan2019automatic,qi2020chest} three different fusion strategies for the optimal fusion of features were investigated. The authors have shown that late-fusion outperformed early and mid-level fusion operations. Therefore we adopt the late-fusion strategy during this work. In our design, predictions are made based on high-level features from both network streams (Fig.\ref{pipeline}-Step1 and 4). During this work we utilize ResNet50 \cite{he2016deep} as the encoder network for both streams. Prior SSL methods used AlexNet \cite{gyawali2020semi} as their main architecture, however, in \cite{qi2020chest} ResNet50 outperformed the pre-trained AlexNet, for classifying COVID-19 from $CXR(x,y)$ images, and had slightly improved mean accuracy compared to  SonoNet64 \cite{baumgartner2017sononet}, XNet(Xception)\cite{chollet2017xception}, InceptionV4(Inception-Resnet-V2)\cite{szegedy2016inceptionv4} and EfficientNetB4 \cite{tan2019efficientnet}.

%In Step 1, a multi-feature network architecture consisting of two mono-feature networks with late fusion for processing both original and enhanced CXR scans was used as a teacher model corresponding to $ T $, due to its outstanding performance in classification of CXR scans \cite{qi2020chest}. 

In our proposed work we denote $x_{ld}$ as labeled data and $y_{ld}$ the corresponding label. Unlabeled data is denoted as $ x_{uld}$ and the generated pseudo label is denoted as $\hat{y}_{uld}$. The teacher model, denoted as $T$, after finetuning with the labeled data, $x_{ld}$, performs a forward run on unlabeled images $ x_{uld}$ to obtain the class distribution $ P(.|x_{uld};\theta_T) $, where $ \theta_T $ denotes the parameters of $ T $.

From this distribution, the trained teacher model predicts the pseudo-label  $\hat{y}_{uld}$ for each image according to softmax prediction vector. 

Once the teacher model, $T$, generates pseudo labels from the unlabeled data, the top $ K $ percent of images in each class, are retained as new positive training samples for each class. The ranking is done based on the corresponding label classification score. Optimization of hyperparameter $ K $ is performed using 10\% of labeled and validation data. In terms of the selection of $ K $, a small $ K $ provides the simple and clean images without much labeling noise for each class. When $ K $ increases the images are less obvious and noisier introducing lots of false positives. Therefore, there is a significant trade-off on $ K $. Based on our experiment, the classification accuracy goes down significantly when $ K>0.35 $. And $ K=0.25 $ gives us the best performance and it corresponds to 75\% of the available unlabeled training dataset.

The paired $ x_{uld}, \hat{y}_{uld} $ is then shown to the student model ,denoted as $ S $, to optimize its parameter $\theta_{S}$. The optimization is based on the gradient calculated by back-propagation from the cross-entropy loss using SGD optimizer:

\begin{equation}
	\theta^{(t+1)}_{S}:=\theta^{t}_{S}-\eta*\frac{\partial\ell(x_{uld},\hat{y}_{uld};\theta_{S})}{\partial\theta_{S}}\bigg|_{\theta_{S}=\theta_{S}^{(t)}}=\theta_{S}^{(t)}-\eta*g_{S}^{(t)},
\end{equation}
where $ \eta $ is the learning rate and $ \ell(x,y;\theta) $ denotes the cross entropy loss calculated between input $ x_{uld} $ and label $ \hat{y}_{uld} $ with parameter $ \theta $

After the student model optimizes its parameters, using Equation 1, its parameters $ \theta_{S}^{(t+1)} $ is further optimized on a labeled sample $ x_{ld}, y_{ld} $ using cross entropy again to minimize the loss:
\begin{equation}
	\min_{\theta_{S}^{(t+1)}}\ell(x_{ld}, y_{ld}; \theta_{S}^{(t+1)})
\end{equation}
Clearly, $ \ell(x_{ld}, y_{ld}; \theta_{S}^{(t+1)}) $ depends on $ \theta_{S}^{(t+1)} $, which in turn depends on the pseudo label $ \hat{y}_{uld} $
%In the Step 4, the student model learns from the optimized pseudo-labeled. The update is based on the gradient calculated by back-propagation from cross entropy loss. 
%
%\sim P(.|x_{unl};\theta_T)
%the pesudo-label, $\hat{y}_{unl}\sim P(.|x_{unl};\theta_T)$,  was provided by \textit{T} to create pseudo-labeled dataset. . FirstThe pesudo-labeled dataset can profitably be filtered by using P and K parameters. First, w
%
%,  on each sample to create pseudo-labeled data. The pseudo-labeled dataset can profitably be filtered by using P and K parameters. First after softmax

\noindent\textbf{Dataset:} We evaluate the performance of proposed method on following datasets: BIMCV \cite{de2020bimcv}, COVIDx \cite{wang2020covid}, COVID-19-AR\cite{Desai2020}, and MIDRC-RICORD-1c\cite{doi:10.1148/radiol.2021203957}. The COVID-19-AR and MIDRC-RICORD-1c were downloaded from the Cancer Imaging Archive (TCIA) Public Access\cite{TCIA} thus denoted as TCIA dataset. The total data collection is consist of 18,691 CXR scans from 16,817 subjects. The detailed data distribution is shown in Table \ref{table:all} and 
%\begin{figure}
%	\begin{center}
%	\includegraphics[width=9cm]{data_fig2.png}
%	\caption{Class distribution of evaluation dataset.}
%	\label{fig:data}
%\end{center}
%\end{figure}

\begin{table}
	\caption{Data distribution of total CXR scans}
	\label{table:all}
	\renewcommand{\arraystretch}{1.2} 
	\centering
	\begin{tabular}{|c|c|c|c|c|c|} 
		%		\hline
		%		\multicolumn{5}{|c|}{Dataset-1}     \\ 
		\hline
		&\multirow{2}{*}{Normal\cite{wang2020covid}} & \multirow{2}{*}{Pneumonia\cite{wang2020covid}} & \multirow{2}{*}{COVID-19\cite{wang2020covid}} & \multirow{2}{*}{COVID-19\cite{de2020bimcv}} & \multirow{2}{*}{COVID-19\cite{TCIA}}\\
		& (COVIDx)&(COVIDx) & (COVIDx) & (BIMCV) & (TCIA) \\ 
		\hline
		\# images   & 8851  & 6045     & 400 & 2167 & 1228\\ 
		\hline
		\# subjects & 8851   & 6031     & 301 & 1183 & 451\\                            
		\hline
	\end{tabular}
\end{table}

\begin{table}
	\caption{Class distribution of each dataset}
	\label{cross-val}
	\centering
	\renewcommand{\arraystretch}{1.2} 
	\begin{tabular}{|c|c|c|c|c|} 
		%		\hline
		%		\multicolumn{5}{|c|}{Dataset-1}     \\ 
		\hline
		&             & Normal & Pneumonia & COVID-19   \\ 
		\hline
		\multirow{2}{2.5cm}{Training \& Stop \& Val data}   & \# images   & 1811  & 1811     & 1811  \\ 
		\cline{2-5}
		& \# subjects & 1811   & 1810    & 1038    \\ 
		\hline
		\multirow{2}{*}{Test-1 data} & \# images   & 756   & 756      & 756  \\ 
		\cline{2-5}
		& \# subjects & 756   & 756      & 446   \\ 
		\hline
		\multirow{2}{*}{Test-2 data} & \# images   & 6284  & 3478     & 1228   \\ 
		\cline{2-5}
		& \# subjects & 6284  & 3465     & 451   \\                              
		\hline
	\end{tabular}
\end{table}

%The COVID-19-AR and MIDRC-RICORD-1c were downloaded from the Cancer Imaging Archive (TCIA) Public Access\cite{TCIA} thus called TCIA dataset. 

A subset containing 5,433 with a balanced distribution of classes consisting of 1- all COVID-19 images from COVIDx, 2- partial normal and pneumonia images from COVIDx, and 3- partial COVID-19 images from BIMCV were randomly selected as our evaluation dataset (Table. \ref{cross-val}). This data was split into a training dataset (4,400 scans), validation dataset (544 scans), and early stopping set (489 scans). We keep our validation set to a realistic size as heavy hyperparameter tuning on a large validation set may have limited real-world applicability \cite{oliver2018realistic}. All the random splits were repeated five times and average results are reported. During the random split, the evaluation and testing data did not include the same patient scans. We generate small labeled data by using 10\%, 20\%, and 30\% of the training dataset and treat the rest of the training data as unlabeled examples. We would like to reiterate that 10\%, 20\%, and 30\% of our training data corresponds to 2.35\%, 4.7\%, and 7.06\% of the full labeled data (18,691 CXR scans). 

Two test datasets were generated to show the robustness of our proposed approach. The remaining COVID-19 images in BIMCV plus same amount of images in normal and pneumonia classes from COVIDx form our Test-1 dataset. All remaining images in COVIDx and all images in TCIA dataset were used to form our Test-2 dataset. The joined test data corresponds to 70.93\% of the full dataset (18,691 CXR scans). 

To evaluate the performance of our multi-feature SSL using $CXR(x,y)$ and $ MF(x,y) $ simultaneously to guide both teacher and student models, denoted as MF-TS, we compared it against three types of SSL methods: 1- SSL guided by $CXR(x,y)$, denoted as CXR-TS (this is also the baseline for Billion Scale SSL \cite{DBLP:journals/corr/abs-1905-00546}); 2- SSL guided by $MF(x,y)$, denoted as Enh-TS; 3- SSL with the teacher model guided by both $ CXR(x,y)$ and $MF(x,y)$ and the student model guided by $CXR(x,y)$ only, denoted as MF-T. We also compare our method against previously developed SSL methods: Temporal Ensembling \cite{laine2016temporal}, and Pseudo Labeling \cite{lee2013pseudo}. In addition, comparison between our proposed SSL method and supersized learning (SL) is also conducted.Three leading network architectures are incorporated: ResNet50\cite{he2016deep}, which trained by varying percent of labeled CXR samples (10\%, 20\%, 30\%, and 100\% of training data), XNet(Xception)\cite{chollet2017xception} and InceptionV4(Inception-Resnet-V2)\cite{szegedy2016inceptionv4}, which are trained on 100\% of the labeled training data only. During comparative evaluation, we use the same amount of validation samples to tune hyperparameters for the investigated baselines. We evaluate the proposed and all the baseline methods by reporting the mean accuracy, precision, recall, and F1-score values.

\section{Results}
%We validate our proposed technique from following perspectives. First, we compared our method with the state of the arts supervised techniques including ResNet50\cite{he2016deep}, InceptionV4\cite{szegedy2016inceptionv4}, and EfficientNetB4\cite{tan2019efficientnet}. In addition, we performed a data ablation study of our SSL method by using 10\%, 20\%, 30\%, and 70\% of labels. Finally, for the sake of fairness, we also evaluated the performance of our SSL approach to some other the state of the art SSL methods. 

All proposed networks were trained for 50 epochs, using the early stopping technique \cite{DBLP:journals/corr/MahsereciBLH17} to avoid overfitting, a learning rate of 0.001 for the first epoch and a learning rate decay of 0.1 every 15 epochs with a mini-batches of size 32. All images were normalized to have zero mean and unit variance and resized to the suitable size for each network during training. All techniques were implemented in Python using the Pytorch framework

Firstly, we notice that all the investigated methods have better overall performance across all metrics for Test-1 dataset compared to Test-2 dataset (Table.\ref{tab:quant}). Test-2 is a more challenging dataset, which has a significantly larger number of samples compared with Test-1 and the cases of COVID-19 in Test-2 from different populations and regions. 

%\begin{figure}
%	\centering
%	\includegraphics[width=9cm]{quant_res4.png}
%	\caption{Quantitative results obtained from Test-1 and Test-2 data. Green shaded region corresponds to the highest scores obtained. 10\%, 20\%, and 30\% of our labeled training data corresponds to 2.35\%, 4.7\%, and 7.06\% of the full labeled data (18,691 CXR scans).}
%	\label{fig:quant}
%\end{figure}

\begin{table}[]
	\begin{center}
		\caption{Quantitative results obtained from Test-1 and Test-2 data. Green shaded region corresponds to the highest scores obtained. 10\%, 20\%, and 30\% of our labeled training data corresponds to 2.35\%, 4.7\%, and 7.06\% of the full labeled data (18,691 CXR scans).}		
		\label{tab:quant}
		\begin{tabular}{c c cc cc cc cc  }
			\toprule
			%			\hline 
			\multirow{2}{*}{\textbf{Method}} & \multirow{2}{2cm}{\textbf{Labeled Sample(\%)}}  & \multicolumn{2}{c}{\textbf{ Precision }} & 
			\multicolumn{2}{c}{\textbf{ Recall }} & \multicolumn{2}{c}{\textbf{ F1-Scores }} &\multicolumn{2}{c}{\textbf{ Top-1(\%)}}\\
			\cmidrule(r){3-4}\cmidrule(lr){5-6}\cmidrule(lr){7-8}\cmidrule(l){9-10}
			&   & Test-1  & Test-2 & Test-1 & Test-2 & Test-1& Test-2&Test-1&Test-2\\
			\hline
			XNet\cite{chollet2017xception}                & 100 & 0.93& 0.91& 0.93& 0.93& 0.93& 0.92& 93.02& 91.69\\			
			\hline
			InceptionV4\cite{szegedy2016inceptionv4}      & 100 & 0.93& 0.90& 0.93& 0.93& 0.93& 0.92& 92.72& 91.82\\
			\hline
			\multirow{4}{*}{ResNet50\cite{he2016deep}}    &10 & 0.86& 0.81& 0.86& 0.86& 0.86& 0.83& 86.46& 84.29\\
			&20 & 0.90& 0.84& 0.90& 0.89& 0.90& 0.86& 89.61& 87.29\\
			&30 & 0.91& 0.88& 0.91& 0.92& 0.91& 0.90& 91.01& 90.19\\
			&100& 0.94& 0.92& \cellcolor{green!25}\textbf{0.94}& \cellcolor{green!25}\textbf{0.94}& \cellcolor{green!25}\textbf{0.94}& \cellcolor{green!25}\textbf{0.93}& 93.54& 92.41\\
			\hline
			\multirow{3}{2cm}{Temporal Ensembling\cite{laine2017temporal}}  &10& 0.76& 0.58& 0.74& 0.57& 0.74& 0.57& 74.00& 66.50\\
			&20& 0.83& 0.71& 0.82& 0.67& 0.82& 0.68& 81.85& 77.79\\
			&30& 0.86& 0.68& 0.85& 0.65& 0.85& 0.66& 85.23& 78.47\\
			\hline
			\multirow{3}{*}{Pseudo-Labeling\cite{arazo2020pseudolabeling}}  & 10& 0.88& 0.57& 0.88& 0.57& 0.88& 0.55& 88.36& 74.12\\
			& 20& 0.92& 0.84& 0.92& 0.77& 0.92& 0.79& 92.06& 83.72\\
			& 30& 0.93& 0.90& 0.93& 0.92& 0.93& 0.91& 93.17& 90.02\\
			
			\hline
			&10& 0.90& 0.88& 0.90& 0.92& 0.90& 0.90& 90.29& 89.43\\
			CXR-TS													&20& 0.92& 0.90& 0.92& 0.93& 0.92& 0.91& 91.59& 90.69\\
			(Billion-Scale\cite{DBLP:journals/corr/abs-1905-00546})	&30& 0.92& 0.91& 0.92& 0.93& 0.92& 0.92& 92.19& 91.52\\
			\hline
			\multirow{3}{*}{Enh-TS}            				&10& 0.91& 0.85& 0.91& 0.81& 0.91& 0.82& 90.99& 85.98\\
			&20& 0.92& 0.90& 0.92& 0.89& 0.92& 0.89& 92.28& 89.70\\
			&30& 0.92& 0.90& 0.92& 0.90& 0.92& 0.90& 92.47& 90.17\\
			\hline
			\multirow{3}{*}{MF-T}               &10 & 0.91& 0.90& 0.91& 0.92& 0.91& 0.91& 91.46& 90.26\\
			&20 & 0.92& 0.90& 0.92& 0.93& 0.92& 0.91& 92.25& 91.01\\
			&30 & 0.93& 0.91& 0.92& 0.93& 0.92& 0.92& 92.49& 91.79\\
			\hline
			\multirow{3}{*}{MF-TS}              & 10& 0.92& 0.90& 0.92& 0.91& 0.92& 0.90& 92.43& 90.08\\
			& 20& 0.93& 0.92& 0.93& 0.93& 0.93& 0.93& 93.41& 91.84\\
			& 30& \cellcolor{green!25}\textbf{0.94}& \cellcolor{green!25}\textbf{0.93}& \cellcolor{green!25}\textbf{0.94}& \cellcolor{green!25}\textbf{0.94}& \cellcolor{green!25}\textbf{0.94}& \cellcolor{green!25}\textbf{0.93}& \cellcolor{green!25}\textbf{93.61}& \cellcolor{green!25}\textbf{92.47}\\
			%			\hline
			\bottomrule
		\end{tabular}
	\end{center}
\end{table}

The proposed MF-TS model achieves an equivalent performance, by only using 30\% labeled training data, in comparison with ResNet50 \cite{he2016deep} and outperforms the XNet \cite{chollet2017xception} and the InceptionV4 \cite{szegedy2016inceptionv4} for both Test-1 and Test-2 data (Table.\ref{tab:quant}). For Test-1 data, ResNet50 \cite{he2016deep} accuracy drops down to 86.46\% when trained on 10\% labeled training data while the proposed MF-TS achieves a significantly higher mean accuracy of 92.43\% ($p<0.05$ with paired t-test). For Test-2 data we observe a similar significant ($p<0.05$ with paired t-test) improvement between ResNet50\cite{he2016deep} and our proposed MF-TS network architecture (84.29\% vs 90.08\% mean accuracy). We also compared our multi-feature guided method with three SSL techniques. From Table.\ref{tab:quant}, we observe that MF-TS offers a substantial improvement over Temporal Ensembling \cite{laine2017temporal} for all metrics at every different labeled sample. The largest differences are 18.43\% and 23.58\%, at 10\% labeled training data, in terms of overall mean accuracy for Test-1 and Test-2 data respectively. Pseudo-labeling \cite{arazo2020pseudolabeling} provides close results for all metrics compared with the MF-TS for the Test-1 dataset. Our method, however, provides better stability in testing different datasets. In terms of overall accuracy, the average change, across different labeling samples between Test-1 and Test-2, of MF-TS is 1.69\% against 8.58\% of Pseudo-labeling \cite{arazo2020pseudolabeling}. Our MF-TS model, as an extension of CXR-TS (Billion-Scale \cite{DBLP:journals/corr/abs-1905-00546}), improves the performance from all aspects. MF-TS model achieves significantly improved mean accuracy compared to CXR-TS model in Test-1 and Test-2 (paired t-test $p<0.05$) except for 10\% labeled training data results of Test-2 data (Table.\ref{tab:quant}).

To further support the assertion that features from enhanced images are beneficial for SSL, we list two additional models, which are Enh-TS and MF-T. Enh-TS and MF-T slightly outperform CXR-TS as shown in Table.\ref{tab:quant}. Investigating Table.\ref{tab:quant}, only MF-T has a better performance in all metrics compared with CXR-TS. The improvement mainly comes from pseudo-label generation. Our fusion model has a more precise prediction for unlabeled samples, and thus it is beneficial for training a new student model.

\section{Conclusions}
We presented a novel multi-feature SSL method for classifying COVID-19 disease from CXR scans. Most prior work that evaluates SSL methods uses a subset of labeled data and large unlabeled data (70\%) for training and a small subset for testing (20\%). To simulate a more realistic scenario, which is often found in medical imaging applications where both obtaining data and labeling efforts are expensive, we opted to use a relatively small labeled dataset (16.48\%) and validation dataset (5.53\%)for training. We exhibit, on a small training ($23.54\%$) and large testing dataset ($>70\%$), that the proposed multi-feature SSL provides improved classification results for diagnosing COVID-19 from CXR scans compared to prior SSL and SL methods. Our results suggest the feasibility of using local-phase CXR image features for improving the success rate of SSL methods and provide a strong foundation for future developments. Future work will include more extensive evaluation and investigation of the proposed method for classifying COVID-19 disease from CT and ultrasound data using SSL methods.

%We presented a novel multi-feature SSL method for classifying COVID-19 disease from CXR scans. Most prior work that evaluates SSL methods uses a subset of labeled data and large unlabeled data (70\%) for training and a small subset for testing (20\%). To simulate a more realistic scenario, which is often found in medical imaging applications where both obtaining data and labeling efforts are expensive, we opted to use a relatively small unlabeled dataset (16.48\%) for training. We have also kept the validation dataset size small (5.53\%) to simulate a realistic scenario. We exhibit, on a small training ($23.54\%$) and large testing dataset ($>70\%$), that the proposed multi-feature SSL provides improved classification results for diagnosing COVID-19 from CXR scans compared to prior SSL and SL methods. Our results suggest the feasibility of using local-phase CXR image features for improving the success rate of SSL methods and provide a strong foundation for future developments. Future work will include more extensive evaluation and investigation of the proposed method for classifying COVID-19 disease from CT and ultrasound data using SSL methods.

\newpage
\bibliographystyle{splncs04}
\bibliography{manuscript}

\end{document}